\documentclass[a4paper]{article}
\usepackage{amsmath,graphicx}
\usepackage{amssymb}
\usepackage{url}
\usepackage{algorithmic, algorithm}
\usepackage{graphicx}
\usepackage{caption}
\usepackage{subcaption}
\newcommand\Tau{\mathcal{T}}

\usepackage{mathtools}

\title{Investigating the Potential of Pseudo Quadrature Mirror Filter-Banks in Music Source Separation Tasks}

\author{
	Stylianos Ioannis Mimilakis\\
	Fraunhofer-IDMT, Ilmenau, Germany\\
	\texttt{mis@idmt.fraunhofer.de}
	\and
	Gerald Schuller\\
	Technical University of Ilmenau, Ilmenau, Germany \\
	\texttt{gerald.schuller@tu-ilmenau.de}
	\date{}
}

\begin{document}
\maketitle
\begin{abstract}
Estimating audio and musical signals from single channel mixtures often, if not always, involves a transformation of the mixture signal to the time-frequency (T-F) domain in which a masking operation takes place. Masking is realized as an element-wise multiplication of the mixture signal's T-F representation with a ratio of computed sources' spectrogram. Studies have shown that the performance of the overall source estimation scheme is subject to the sparsity and disjointness properties of a given T-F representation. In this work we investigate the potential of an optimized pseudo quadrature mirror filter-bank (PQMF), as a T-F representation for music source separation tasks. Experimental results, suggest that the PQMF maintains the aforementioned desirable properties and can be regarded as an alternative for representing mixtures of musical signals.
\end{abstract}

\section{Introduction}
\label{sec:intro}
The separation of audio signals from mixtures is an active research area in the field of audio signal processing. The main objective is to estimate individual auditory components from an observed mixture. By doing so, a series of applications can be derived, spanning from assisting music information retrieval systems (MIR) to audio re-purposing tasks, such as spatial up-mixing and music reproduction \cite{vinvent_ica03}.

In relevant literature, each auditory component is indicated as a \textit{source} and the issue of estimating sources within a mixture that convey music information is commonly referred to as \textit{music source separation} \cite{burred_phd}. Research in music source separation has focused in both multi-channel \cite{fitz_projet} and single channel \cite{cano, cauchy_nmf} cases. For the examination of time-frequency representations, the current investigation is constrained to the single channel (\textit{monaural}) case.

The source estimation from monaural mixtures is achieved through \textit{time varying filtering} adapted to the targeted sources. More specifically, the mixture signal is transformed to the T-F domain, often using a short time Fourier transform (STFT). Through an appropriate method, such as the non-negative matrix factorization or a phase structured method \cite{cauchy_nmf, cano}, spectral models of the sources to be separated are derived. Then from a ratio of spectral models, gain functions are computed \cite{ps_masks, liutkus_alpha}. These functions form T-F \textit{masks} which allow the estimation of a single source through an element-wise multiplication of the mixture T-F representation and the masks.

A significant amount of research has been devoted to the development of ideal signal representations, for optimal filtering, de-noising, and source estimation scenarios. Such studies have underlined that signal representations based on STFT, usually suffer from undesired signal energy leakage between neighbouring frequency bins (\textit{sub-bands}).
This is caused by applying a finite length \textit{windowing} function to the discrete time Fourier transform (DTFT) to obtain the STFT, resulting into sub-band filters with wide transition bands which overlap with neighbouring ones.
As a consequence, two important properties \textit{sparsity} and \textit{disjointness} are not fully exploited by representations based on STFT \cite{union_mark, giannoulis_disjointness}.

Sparsity allows a more accurate computation of the contribution of each source in each T-F sample, while disjointness refers to an ideally unique contribution of one source to a single T-F sample. In \cite{fevotte_disjointness} it is shown that over-complete transforms, such as the short-time discrete
cosine transform (DCT) and unions of discrete cosine and wavelet transforms, fail to improve the overall sparsity and separation performance of various types of sources, compared to the modified discrete cosine transform (MDCT). On the other hand, cosine and wavelet packets did not provide a significant improvement over MDCT in evaluation metrics usually employed in source separation performance measurements \cite{vincent_disjointness}. Burred and Sikora \cite{burred_disjointness} examined auditory filter-banks as alternative sparse representations. These included Bark-scaled and equal rectangular bandwidth (ERB) filter-banks that produced sparser representations resulting into better source separation performance. More recently in \cite{giannoulis_disjointness}, transforms such as pitch-synchronous STFT, constant Q transform (CQT), and MDCT are evaluated in terms of disjointness and sparsity, with MDCT providing the best performance.

This work examines the capabilities of a cosine modulated filter-bank, namely pseudo quadrature mirror filter-bank (PQMF), for music source separation tasks. The implementation of the filter-bank is based on the framework of poly-phase matrices presented in \cite{schuller_mpr}. For assessing the performance of the PQMF, subject to music source separation, two objective metrics commonly used in the state of the art were computed: i) w-disjoint orthogonality (WDO) \cite{yilmaz_masks}, measuring the degree of overlap that multiple sources have in a given representation, and ii) sparsity using the Gini index \cite{sparsity_measures}. For comparison, two additional filer-banks namely STFT and MDCT are taken into account, since they are frequently used in music source separation tasks \cite{liutkus_alpha, mitianoudis}. The DSD$100$ audio corpus\footnote{DSD$100$ Dataset: \url{http://liutkus.net/DSD100.zip}} was used for computing the above metrics. It includes $4$ categories of professionally produced music sources, consisting of \textit{bass, drums, singing voice} and \textit{other}.

\section{PQMF Overview}
\label{sec:secII}
The PQMF is a special case of quadrature mirror filter-banks (QMF) with a near-perfect reconstruction property, in which aliasing cancellation takes place
only in adjacent frequency sub-bands \cite{jos_smith}. For sub-bands whose aliasing components are not canceled, 
band-pass filters with maximum attenuation are employed in order to suppress the aliasing components.
Designing such filter-bank consists of constructing two poly-phase matrices $\mathbf{P}^{a}(z)$ and $\mathbf{P} ^{s}(z)$, for the \textit{analysis} and \textit{synthesis} operations respectively. They are expressed in the $z$ domain via $P_{n,k}(z) = \sum_{m=0}^{L-1} P_{n,k}(m) z^{-m}$, with $n$ denoting the rows and $k$ the columns of the matrix, over the time-frames $m$ and overlap $L$. 

In practice, the coefficients of the above mentioned matrices have to be determined such as they approximate the reconstruction property $\mathbf{P}^{s}(z) = \mathbf{P}^{a}(z)^{-1} z^{-d}$ with $d$ being a necessary delay to make the system causal \cite{schuller_ld}. These coefficients are connected to the time-domain samples of a windowing function $h(n)$ \cite{schuller_ld}, which can be computed by means of convex optimization \cite{pqmf_convex}, modulated by cosine basis functions. For purposes of this work, the windowing function was optimized to obtain $N = 1024$ frequency sub-bands using a filter length of $M = 8192$ time-domain samples, which results in an overlap of $L = 8$. An overview of the implementation is given in Algorithm~1.Figures~1a and 1b demonstrate the result of the least squares minimization (opt-PQMF) and its corresponding frequency response compared to broadly used windowing functions, \textit{Hamming} and \textit{Sine} defined as:

\begin{equation}
\begin{split}
w(n)_{hamm} & = 0.54 - 0.46 \cos (\frac{2\pi n}{M - 1}) \\
w(n)_{sine} & = \sin(\frac{\pi}{M}(n+0.5))
\end{split}
\end{equation}
for $n  = 0, _{\cdots}, M - 1 \text{ and } M = 2048.$

\begin{algorithm}
\caption{: PQMF Implementation}
\begin{algorithmic}[1]
 \STATE Randomly initialize a windowing function $h(n)$ of total $M=LN$ samples, where $N$ is the number of frequency sub-bands $k$ and $L$ is the overlap factor.
 
 \STATE Through least squares minimization, approximate the reconstruction condition via: \\ $|H(e^{j\omega})|^2 + |H(e^{j(\pi/N) - \omega})|^2 = 2 \text{, for } 0 < |\omega| < \frac{\pi}{2N}$ \text{ and }
  $|H(e^{j\omega})|^2 = 0 \text{, for } \omega > \frac{\pi}{N}$, where 
   $|H(e^{j\omega})|$ is the DTFT of $h(n)$.
 
  

  \STATE After the optimization the analysis and synthesis polyphase matrices are constructed as follows: \\ ${P^a_{n',k}}(m) = h(mN + n) \sqrt{\frac{2}{N}} \cos(\frac{\pi}{N} (k + \frac{1}{2}) (LN - 1 - mN+n - \frac{N}{2} + 0.5))$\\
                                 
  ${P^s_{k,n}}(m) = h(mN + n) \sqrt{\frac{2}{N}} \cos(\frac{\pi}{N} (k + \frac{1}{2}) (mN + n - \frac{N}{2} + 0.5))$, where $k,m,n \in \mathbf{Z}: 0 \leq m < L$,  $\forall k,n \in \{0,{ }_{\cdots}, N-1\}$, and $n' = N - 1- n$.
  

\STATE For the analysis and synthesis of an input signal $x(n)$, let it be represented by a vector $\mathbf{x}_m(n) \in \mathbf{R}^{N}$ composed by down-sampled elements $\mathbf{x}_m(n) = [x(mN), x(mN + 1), { }_{\cdots}, x(mN + N - 1)]$. By expressing $\mathbf{x}_m(n)$ in the $z$-domain, denoted as $\mathbf{X}(z)$, its approximation by the PQMF filter-bank is given by $\hat{\mathbf{X}}(z) = \mathbf{X}(z) \mathbf{P}^{a}(z) \mathbf{P}^{s}(z) $.
\end{algorithmic}
\end{algorithm}
\begin{figure}[!t]
 \centering
  \begin{subfigure}[b]{1\textwidth}
    \includegraphics[width=0.9\textwidth, height = 0.4\textwidth]{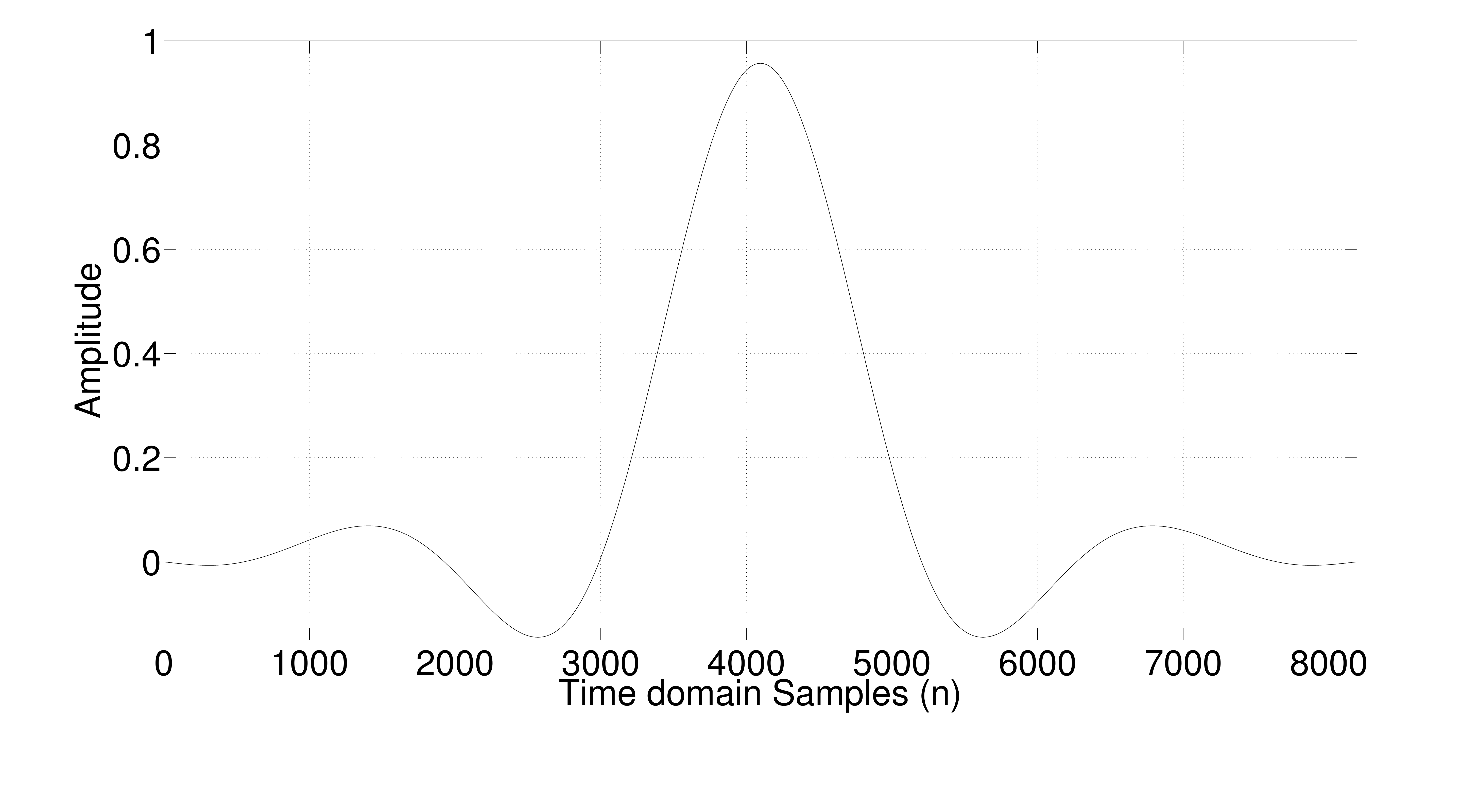}
    \caption{Result from the least-squares minimization.}
    \label{fig:f1}
  \end{subfigure}
 \begin{subfigure}[b]{1\textwidth}
    \includegraphics[width=0.85\textwidth, height = 0.45\textwidth]{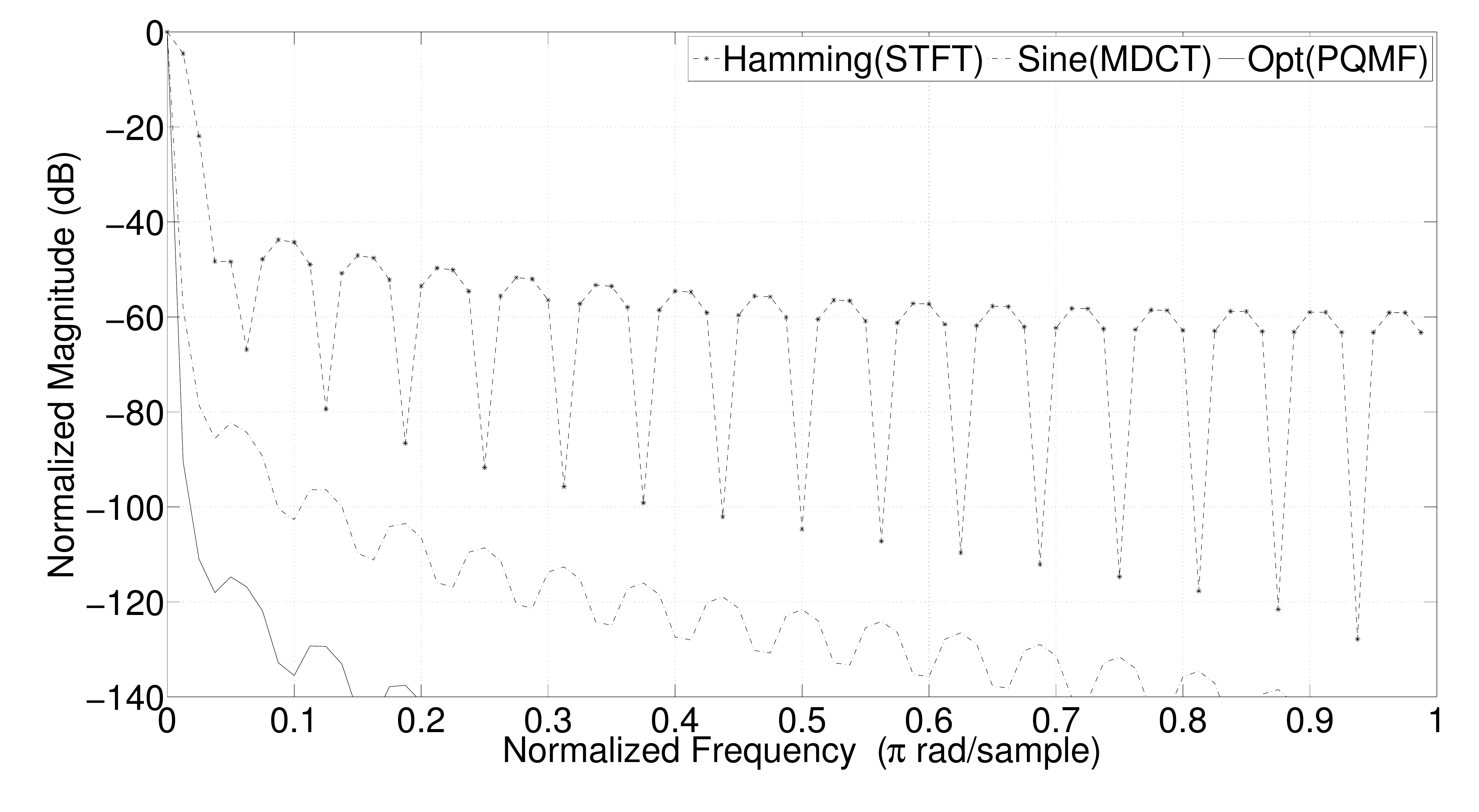}
    \caption{Frequency responses  of three windowing functions, demonstrating the suppression of undesired spectral leakage between neighbouring sub-bands.}
    \label{fig:f1}
  \end{subfigure}
  \centering
 \caption{Result of the optimization and its frequency response compared to common windowing functions.}
\end{figure}
\vfill

\section{Experimental Procedure}
\label{sec:secIII}
\subsection{Measures of Disjointness \& Sparsity}
Let $\mathbf{s}_j$ be the set of $J$ total additive sources contained in a monaural mixture $\mathbf{x}$. The estimation of a source $\mathbf{\hat{s}}_j$ via time-frequency masking is expressed as:
\begin{equation}
\mathbf{\hat{s}}_j = \Tau^{-1}(M_{j}(\Tau(\mathbf{x})))
\end{equation}
where $M_{j}$ is the mask of the target source $j$ to be separated and $\Tau$ is an operator that maps a time domain signal to the time-frequency domain by the analysis filter-bank. The corresponding counterpart is given by  $\Tau ^{-1}$. For computing the mask $M_j$ the same approach as \cite{yilmaz_masks} is followed. For a set of frequency sub-bands $k$, the mask $M_j$ is computed as:
\begin{equation}
M_j(k) = \begin{cases}
1, \text{ if } {|S_j(k)|} \geq {|U(k)|} \\
0,  \text{otherwise}
\end{cases}
\end{equation}
with $U(k)$ being the T-F representation of the sum of the interfering sources and $S_j(k)$ is the T-F representation of the target source to be estimated by the mask. An approximation of the frequently used w-disjoint orthogonality (WDO) is derived from:
\begin{equation}
\text{WDO} = \text{PSR} - \frac{\text{PSR}}{\text{SIR}}
\end{equation}
where PSR and SIR stand for the \textit{preserved signal ratio} and \textit{signal to interference ratio} respectively, defined as:
\begin{equation}
\text{\small{PSR}} = \frac{{\overset{N-1}{\underset{k = 0}{\sum}}}(M_j(k) |S_j(k)|)^2} {\overset{N-1}{\underset{k = 0}{\sum}}|S_j(k)|^2}
\text{, \small{SIR}} = \frac{{\overset{N-1}{\underset{k = 0}{\sum}}} (M_j(k) |S_j(k)|)^2}{{\overset{N-1}{\underset{k = 0}{\sum}}} (M_j(k) |U(k)|)^2}
\end{equation} 
The values of WDO vary from $0$ to $1$, where $1$ implies a perfect separation and recovery of the target source.

For acquiring sparsity measures, the Gini index ($GI$) \cite{sparsity_measures} was utilized as formulated in Eq.~6.
\begin{equation}
GI = \frac{1}{N} + 1 -2 \bigg[\sum_{k=0}^{N - 1} \frac{|X(k)|}{||X||_1}\bigg(\frac{N - k + 0.5}{N}\bigg)\bigg],
\end{equation}
where $|X(k)|$ is the magnitude of the T-F representation of $\mathbf{x}$, but sub-bands $k$ are reordered by magnitude $|X(0)| \leq |X(1)| \leq |X(N-1)|$ in order to be scaled accordingly. This will result into a more intuitive and robust sparsity estimation compared to typical $\ell_1$, $\ell_2$ norms \cite{gini_II}. The values of $GI$ span from $0$ to $1$, where $1$ indicates that the signal has one significant coefficient and thus, is as sparse as possible.
It should be noted that the index indicating the time frames
is omitted for clarity. As far it concerns the computation of $GI$, an average value over time frames is computed.

\subsection{Audio corpus analysis}
In order to assess the performance of the PQMF in source separation tasks the DSD$100$ dataset was employed. It consists of $100$ professionally produced multi tracks of various music genres, sampled at $44.1$kHz. Each multi-track consists of the target sources which are used as side information for computing WDO. 

In more details, for each multi track a monaural version of the $4$ sources was generated by averaging the two available channels. Afterwards, two types of mixture signals are synthesised. One containing all the monaural sources, for computing the sparsity measure, and one containing only the interfering sources $U$ with respect to the target source $s_j$.

For each of the mixture types and sources contained in a multi-track, the following decomposition methods, which are broadly used in music source separation tasks, were considered for the assessment:
\begin{itemize}
	\item  STFT with a hamming windowing function (STFT-Hm), covering $M = 2048$ samples and $80$\% overlap between adjacent frames; heuristic rules producing desirable performance in music source separation tasks \cite{liutkus_alpha}. Since the analysed signals are real valued, their spectra are Hermitian and the redundant information is discarded, resulting into $N = 1024$ frequency sub-bands.
	
	\item MDCT based on type-IV bases and a sine windowing function covering $M = 2048$ samples with $50\%$ percent of overlap between adjacent time frames, producing a total of $N = 1024$ frequency sub-bands \cite{mitianoudis}.
	
	\item The PQMF as described in Algorithm~1, producing total $N = 1024$ frequency sub-bands using $M = 8192$ samples.
	
\end{itemize}

\section{Results \& Discussion}
\label{sec:secIV}
The results from the disjointness and sparsity measures are demonstrated in Figures~2 and ~3, respectively.
The lower and upper quartiles are depicted with the lower and upper horizontal lines in each box. The interquartile lines and points indicate the median and average values respectively, while crosses denote outliers in the observations. For both metrics $1$ denotes the best possible performance.

By observing Figure~2 it can been seen that both MDCT and PQMF outperform the STFT decomposition, in terms of providing a disjoint representation of mixture signals consisting of music sources. This is also reflected by the sparsity measure illustrated in Figure~3. Real valued transformations provide the sparsest representations. This can be explained by their nor-redundancies in representing signals and the employed windowing functions illustrated Figure~1b, where the energy leakage between neighbouring sub-bands is highly suppressed by the windowing functions incorporated in the real valued representations, stressing out the importance of choosing an appropriate windowing function.
\begin{figure}[!h]
	\centering
		\includegraphics[width=0.9\textwidth, height = 0.8\textwidth]{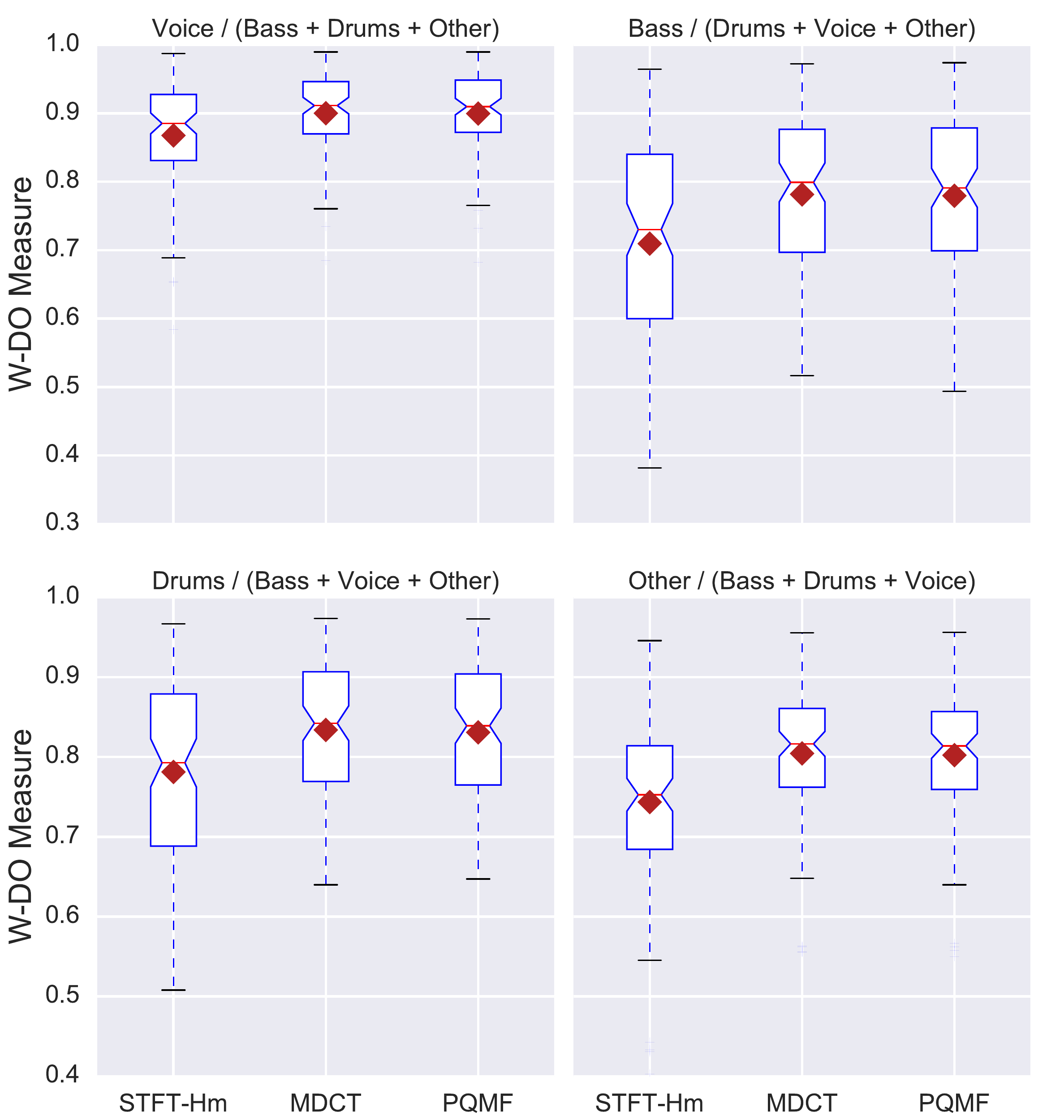}
	\caption{Variation analysis of the disjointness measure from three T-F decompositions, over $4$ categories of music sources.}
\end{figure}
\begin{figure}[!h]
	\centering
	\includegraphics[width=0.85\textwidth, height = 0.65\textwidth]{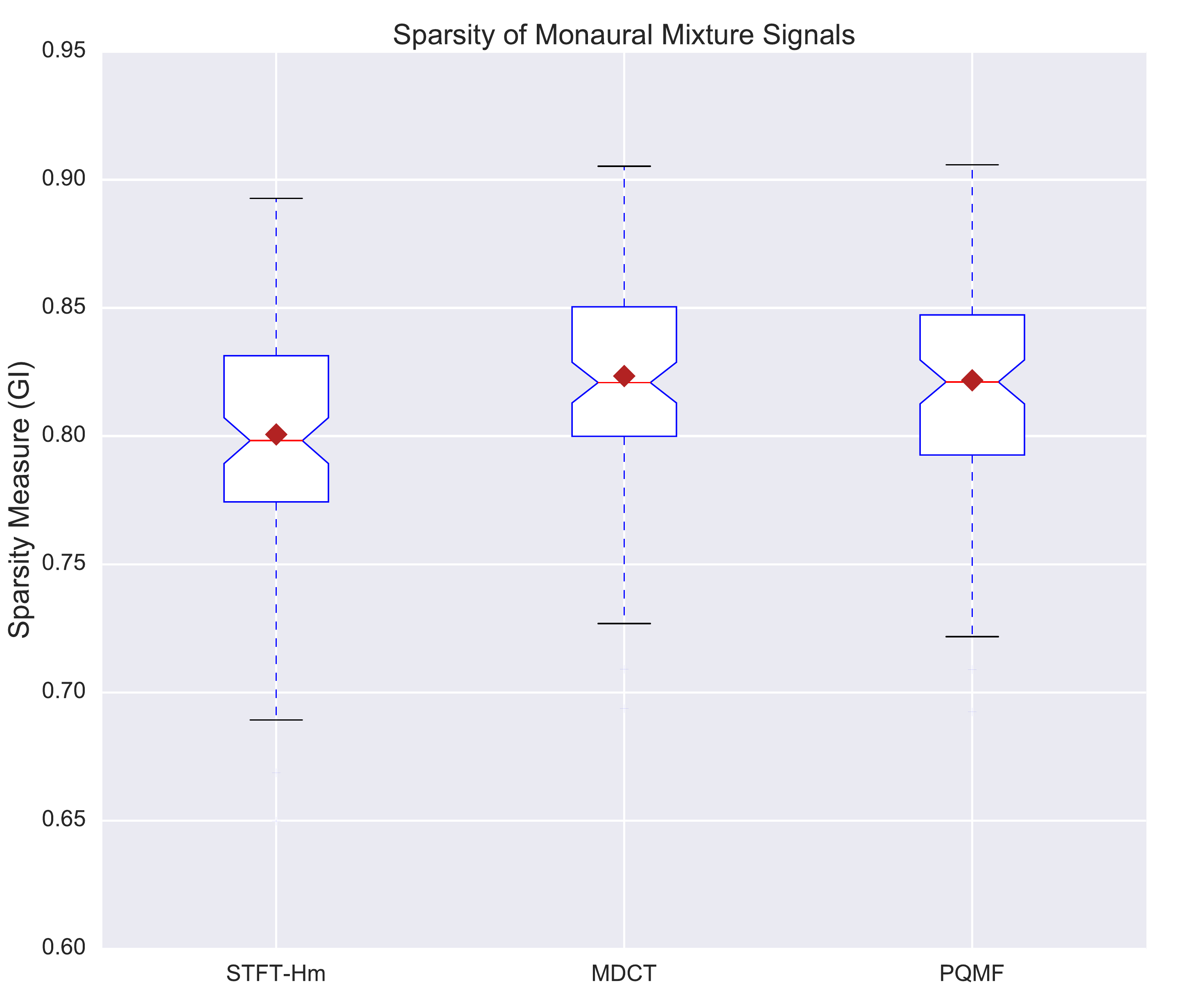}
	\caption{Variation analysis of the Gini index.}
\end{figure}

In general, the overall performance of the PQMF and MDCT is almost identical. Nonetheless, there are some differences to be underlined. Upper quartiles of the disjointness provided by the PQMF are slightly increased for quasi-harmonic harmonic instruments such as voice, contrary to sources having impulsive nature such as drums and other. Additionally, the median values of sparsity measures regarding the PQMF are somewhat higher compared to MDCT, but not for all the mixture signals, since the quartiles of MDCT underline a small gain. These two observations are induced by the
difference in the overlap factors between the MDCT and the PQMF. The increased overlap factor in the PQMF affects the disjointness favouring quasi-harmonic sources, for a small loss of sparsity, which is important for the estimation of impulsive sources.

Since the problem of monaural source separation is summarized as a time-varying filtering process, better leakage suppression in time-frequency representations are emerging \cite{jos_smith, pqmf_convex}, ideally resulting into less \textit{musical distortions}. As Figure~1b points out, such desirable properties can be obtained from a least squares optimization procedure.

\section{Conclusions}
\label{sec:secV}
In this work an optimized pseudo quadrature mirror filter-bank (PQMF) was examined for its performance as a front-end time-frequency decomposition method in music source separation tasks. The PQMF was compared to usual lapped decomposition methods such as STFT and MDCT, which are broadly used for estimating music sources from arbitrary mixtures \cite{fitz_projet, mitianoudis}. The assessment included the following set of metrics: i) W-disjoint orthogonality (W-DO) \cite{yilmaz_masks} and ii) a sparsity measure using Gini index (GI) \cite{sparsity_measures, gini_II}.

Results from an experimental procedure covering professionally produced sources, showed that time-frequency representations derived from cosine modulated filter-banks provide the most disjoint and sparse representations. These two properties are well-acknowledged and desired in music source separation tasks \cite{union_mark}, since they improve the overall performance \cite{fevotte_disjointness}. The filter-bank based on pseudo quadrature-mirror filters provided optimal performance of sparsity and disjointness of quasi-harmonic sources conveying music information and particularly singing voice.

In contrast, MDCT provided the best disjoint representations
for estimating sources with impulsive nature such as drums. The upper and lower quartiles of MDCT denote a small gain of sparsity,
pointing out a relation of sparse representations and the estimation of impulsive sources. Furthermore, a correlation between sparsity, disjointness and windowing functions was also pinpointed. From the perspective of time frequency masking as a filtering operation, optimized windowing functions commonly incorporated in cosine modulated filter-banks, seem to provide fertile representations for processing music signals.
Source code can be found under: \url{https://github.com/TUIlmenauAMS/ASP}

\section{Acknowledgements}
The research leading to these results has received funding from the European Union's H2020 Framework Programme (H2020-MSCA-ITN-2014) under grant agreement no 642685 MacSeNet.

\end{document}